\RequirePackage{amsmath}
\documentclass[pdftex,twocolumn,epjc3]{svjour3}          % twocolumn
\smartqed  % flush right qed marks, e.g. at end of proof
\usepackage[none]{hyphenat}
\usepackage{datetime}
\usepackage{mathtools, cuted, widetext}
\usepackage{amsmath,epsfig,amssymb,multirow}
\RequirePackage{graphicx}
\RequirePackage{mathptmx}      % use Times fonts if available on your TeX system
\RequirePackage{flushend}
\RequirePackage[numbers,sort&compress]{natbib}
\RequirePackage[colorlinks,citecolor=blue,urlcolor=blue,linkcolor=blue]{hyperref}

\journalname{Eur. Phys. J. C}
\def\wtil#1{\widetilde{#1}}
\begin{document}
	\sloppy
	\title{On the choice of beam polarization in  $e^+e^-\to ZZ/Z\gamma$ and 
		anomalous triple gauge-boson couplings }
	%%%%%%%%%%%%%%%%%%%%%%%%%%%%%%%%%%%%%%%%%%%%%%%%%%%%%%%%%%%%%%%%%%%%%%%%%%%%%%
	\author{Rafiqul Rahaman\thanksref{e1,adrss} \and 
		Ritesh K. Singh\thanksref{e2,adrss}}
	\thankstext{e1}{email:rr13rs033@iiserkol.ac.in}
	\thankstext{e2}{email:ritesh.singh@iiserkol.ac.in}
	\institute{Department of Physical Sciences,
		Indian Institute of Science Education and Research Kolkata,
		Mohanpur, 741246, India\label{adrss}}
	
	%\date{Received: \today}
\date{}
\maketitle

\begin{abstract}
The anomalous trilinear gauge couplings of $Z$ and $\gamma$ are studied in
$e^+e^-\to ZZ/Z\gamma$ with longitudinal beam polarizations using a complete 
set of polarization asymmetries for the $Z$ boson. We quantify the goodness of the
beam polarization in terms of the likelihood and find the best-choice of $e^-$ and
$e^+$ polarizations to be ($+0.16$, $-0.16$), ($+0.09$, $-0.10$) and 
($+0.12$, $-0.12$) for $ZZ$, $Z\gamma$ and combined processes, respectively. 
Simultaneous limits on anomalous couplings are obtained for these choices of 
beam polarizations using Markov-Chain--Monte-Carlo (MCMC) for an $e^+e^-$ 
collider running at $\sqrt{s}=500$ GeV and  $\mathcal{L}=100$ fb$^{-1}$. We 
find the simultaneous limits for these beam polarizations to be comparable 
with each other and also comparable with the unpolarized beam case.
\end{abstract}

\section{Introduction}
The gauge boson sector in the Standard Model (SM) remains uncharted even after the 
discovery of Higgs boson~\cite{Chatrchyan:2012xdj} at LHC. Of the gauge-boson self 
couplings (trilinear and quartic), the gauge boson couplings to Higgs, the Higgs self 
couplings, which are the key to understand  Electro Weak Symmetry 
Breaking (EWSB), there have no precise measurements and they need serious attention. 
The future International Linear Collider (ILC)~\cite{Djouadi:2007ik,Baer:2013cma,
Behnke:2013xla} will be a precision testing machine~\cite{MoortgatPick:2005cw} 
which will have the possibility of  polarized initial beams. Two types of 
polarization,  namely longitudinal and transverse, for both initial beams 
($e^-$ and $e^+$) will play an important role in precise measurement of various 
parameters, like the coupling among gauge bosons, Higgs coupling to the top quark, and Higgs 
coupling to the gauge boson. Beam polarization has the ability to enhance the 
relevant signal to background ratio along with the sensitivity of 
observables~\cite{MoortgatPick:2005cw,Andreev:2012cj,Ananthanarayan:2010bt,
Osland:2009dp,Pankov:2005kd}.  It can also be used to separate 
CP-violating couplings from a CP-conversing one~\cite{MoortgatPick:2005cw,
Kittel:2011rk,Dreiner:2010ib,Bartl:2007qy,Rao:2006hn,Bartl:2005uh,Czyz:1988yt,
Choudhury:1994nt,Ananthanarayan:2004eb,Ananthanarayan:2011fr,Ananthanarayan:2003wi} if CP-violation 
is present in  Nature. These potentials of the beam polarizations have been 
explored, for example, to study $\tau$ polarization~\cite{Dreiner:2010ib}, top 
quark polarization~\cite{Groote:2010zf} and its anomalous
 couplings~\cite{Amjad:2015mma}, littlest 
Higgs model~\cite{Ananthanarayan:2009dw}, $WWV$ 
couplings~\cite{Ananthanarayan:2011ga,Andreev:2012cj,Ananthanarayan:2010bt}, 
Higgs couplings to gauge bosons~\cite{Kumar:2015eea,Rindani:2010pi,
Biswal:2009ar,Rindani:2009pb}. 

Here we use  beam polarizations (longitudinal only)  to study anomalous 
trilinear gauge-boson self couplings in the neutral sector  using the complete
set of polarization observables of the $Z$ boson~\cite{Rahaman:2016pqj,
Boudjema:2009fz,Aguilar-Saavedra:2015yza} in the process 
$e^+e^-\to ZZ/Z\gamma$. The anomalous couplings among the neutral gauge boson have 
been studied earlier with unpolarized beam in~\cite{Boudjema:92,Baur:1992cd,
Ellison:1998uy,Baur:2000ae,Ananthanarayan:2005ib,Aihara:1995iq,Gounaris:2000dn,Poulose:1998sd,Senol:2013ym} as 
well as with polarized beams in~\cite{Ots:2006dv,Czyz:1988yt,Choudhury:1994nt,
Ananthanarayan:2004eb,Ananthanarayan:2014sea,
Gounaris:1999kf,Choi:1994nv,Rizzo:1999xj,Atag:2003wm,Ananthanarayan:2003wi}. 
Some of these studies have used  given beam polarizations to enhance the 
sensitivity of observables, while others have used two different sets of beam 
polarizations to construct the observables. We follow the former method and quantify
the likelihood-based goodness of the choice of beam polarizations.

For the process of interest  the anomalous triple gauge-boson couplings is 
given by the Lagrangian~\cite{Gounaris:1999kf,Rahaman:2016pqj}
\begin{eqnarray}\label{eq:aTGC_Lagrangian}
{\cal L}=
\frac{g_e}{M_Z^2} \Bigg [
[f_4^Z (\partial_\mu Z^{\mu \beta}) 
-f_4^\gamma (\partial_\mu F^{\mu \beta})] Z_\alpha 
( \partial^\alpha Z_\beta)\nonumber\\
+[f_5^\gamma (\partial^\sigma F_{\sigma \mu})+
f_5^Z(\partial^\sigma Z_{\sigma \mu}) ] \wtil{Z}^{\mu \beta}Z_\beta\nonumber\\
-  [h_1^\gamma (\partial^\sigma F_{\sigma \mu})
+h_1^Z (\partial^\sigma Z_{\sigma \mu})] Z_\beta F^{\mu \beta}\nonumber\\
-[h_3^\gamma  (\partial_\sigma F^{\sigma \rho})
+ h_3^Z  (\partial_\sigma Z^{\sigma \rho})] Z^\alpha
\wtil{F}_{\rho \alpha}
\Bigg ]. 
\end{eqnarray} 
The coupling $f_i^V$s appear in the $ZZ$ process while $h_i^V$s appear in 
$Z\gamma$ process. Among these couplings $f_4^V$ and $h_1^V$ are $CP$-odd 
while others are $CP$-even. The best limits on these anomalous couplings coming
from LHC are
$|f_i^V| \sim 3\times10^{-3}$~\cite{Khachatryan:2015pba} and 
$|h_i^V| \sim 9\times10^{-4}$~\cite{Aad:2016sau}.
 
The rest of the  paper is organized as follows. 
In~\autoref{sec:beampol_and_polarizaiton_observables} we discuss basic 
formulations and the $Z$ boson polarization observables.
In~\autoref{sec:sensitivity_likelihood} we study beam polarization dependence of
sensitivity and likelihood. We define a \textit{measure of goodness} for the 
choice of beam polarizations and study $ZZ/Z\gamma$ processes to obtain the best
choices. The simultaneous limits are presented for a set of beam
polarizations. 
We conclude in \autoref{sec:conclusion}.

%%%%%%%%%%%%%%%%%%%%%%%%%%%%%%%%%%%%%%%%%%%%%%%%%%%%%%%%%%%%%%%%%%%%%%%%%%%%%%
\section{Beam polarization and polarization observables}
\label{sec:beampol_and_polarizaiton_observables}
%----------------------------------------------------------------------------

The polarization density matrices for $e^-$ and $e^+$ beams are given by
\begin{eqnarray}\label{eq:electron_pol_matrix}
P_{e^-}(\lambda_{e^-},\lambda_{e^-}^\prime)=
\frac{1}{2}\left[
\begin{tabular}{cc}
$(1+\eta_3)$&$\eta_T$\\
$\eta_T$&$(1-\eta_3)$
\end{tabular}
\right] \hspace{0.5cm}\textrm{and}
\end{eqnarray}
\begin{eqnarray}\label{eq:positron_pol_matrix}
P_{e^+}(\lambda_{e^+},\lambda_{e^+}^\prime)=
\frac{1}{2}\left[
\begin{tabular}{cc}
$(1+\xi_3)$&$\xi_Te^{-i\delta}$\\
$\xi_Te^{i\delta}$&$(1-\xi_3)$
\end{tabular}
\right],
\end{eqnarray}
where $\eta_3$ and $\eta_T$ ($\xi_3$ and $\xi_T$) are longitudinal and 
transverse polarization of $e^-$ ($e^+$)  with $\delta$ being the azimuthal 
angle between two transverse polarizations. The positive $x$-axis is taken 
along the transverse polarization of $e^-$ and positive $z$-axis along its
momentum.

%%%%%%%%%%%%%%%%%%%%%%%%%%%%%%%
\begin{figure}
\includegraphics[scale=1]{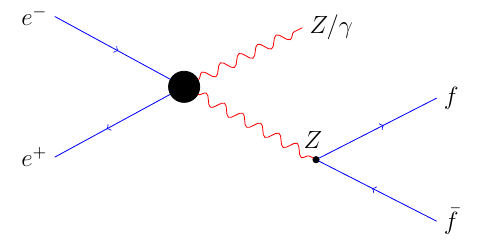}
\caption{\label{fig:Z_production_decay} Feynman diagram  
for production of $Z$ boson and its decay to a pair of 
fermions} 
\end{figure}
%%%%%%%%%%%%%%%%%%%%%%%%%%%%%%%
The density matrix for the production of $Z$ boson in the above process 
(Fig.~\ref{fig:Z_production_decay} ) would be
\begin{eqnarray}\label{eq:density_matrix_beam_pol}
\hspace{-2cm} \rho(\lambda_Z,\lambda_Z^\prime)=
{\cal\:\sum}_{\lambda_{e^-},\lambda_{e^-}^\prime,\lambda_{e^+},\lambda_{e^+}^\prime} 
{\cal M}^\dagger(\lambda_{e^-}^\prime,\lambda_{e^+}^\prime,\lambda_Z^\prime)\times\nonumber\\
{\cal M}(\lambda_{e^-},\lambda_{e^+},\lambda_Z) \times
P_{e^-}(\lambda_{e^-},\lambda_{e^-}^\prime)\times
P_{e^+}(\lambda_{e^+},\lambda_{e^+}^\prime).
\end{eqnarray}
We note that the different helicities can take the following values:
\begin{eqnarray}
\lambda_Z,\lambda_Z^\prime\in\{-1,0,1\} \ \text{and} \ 
\lambda_{e^\pm},\lambda_{e^\pm}^\prime\in\{-1,1\}. 
\end{eqnarray}
For the present work we  restrict ourselves  only to the longitudinal beam 
polarizations, i.e. $\eta_T=0=\xi_T$. With the chosen beam polarizations we 
construct the complete set of eight polarization observables for the $Z$ boson 
along with total cross section in the processes $e^+e^-\to ZZ/Z\gamma$. These 
polarization observables can be obtained analytically from the production 
process as well as from asymmetries constructed from decay distribution of the
particle. The polarization observables consist of a $3$ component vector 
polarization $\vec{P}\equiv(P_x,P_y,P_z)$ and a traceless symmetric rank-$2$ 
tensor  $T_{ij}(i,j=x,y,z)$ with $5$ independent component $T_{xy}$, $T_{xz}$, 
$T_{yz}$, $T_{xx}-T_{yy}$ and $T_{zz }$. The asymmetries in the collider or in 
a Monte Carlo event generator corresponding to $P_i$'s and $T_{ij}$'s are 
$\{A_x, A_y, A_z\}$ and $\{A_{xy}, A_{xz}, A_{yz}, A_{x^2-y^2}, A_{zz}\}$, 
respectively. The asymmetries $A_z$, $A_{xz}$, $A_{yz}$ are zero
in SM and even with polarized beam in both  processes owing to the 
forward-backward symmetry of produced $Z$ in these processes. To make these asymmetries
non-zero we redefine the polarization observables ${\cal O}\in\{P_z,T_{xz},T_{yz}\}$ 
as

\begin{eqnarray}
{\cal O}\to \wtil{{\cal O}}= \frac{1}{\sigma_{Z}} \Bigg[\int^{c_{\theta_0}}_{0} 
{\rm Comb}({\cal O},\rho(\lambda,\lambda')) dc_{\theta_Z}\nonumber\\
-\int^{0}_{-c_{\theta_0}}
{\rm Comb}({\cal O},\rho(\lambda,\lambda')) dc_{\theta_Z}\Bigg],
\end{eqnarray}
 
where $c_{\theta_0}$ is the beam pipe cut and
${\rm Comb}({\cal O},\sigma(\lambda,\lambda'))$ is the combination of production
density matrix corresponding the polarization observable ${\cal O}$ (see 
Ref.~\cite{Rahaman:2016pqj}). For example, 
with ${\cal O}=P_z$ one has 
$${\rm Comb}(P_z,\rho(\lambda,\lambda'))=\rho(+1,+1) -\rho(-1,-1)$$
and the corresponding modified polarization is given by
\begin{eqnarray}
\wtil{P}_z=\frac{1}{\sigma_{Z}} \bigg[\int^{c_{\theta_0}}_{0}    
\bigg[\rho(+1,+1) -\rho(-1,-1) \bigg] dc_{\theta_Z}\nonumber\\
 -\int^{0}_{-c_{\theta_0}}\bigg[\rho(+1,+1) -\rho(-1,-1) \bigg]
dc_{\theta_Z}\bigg]. 
\end{eqnarray} 
The asymmetries $\wtil{A}_z$ corresponding to the modified polarization 
$\wtil{P}_z$ is given by:
\begin{eqnarray}
\wtil{A}_z\equiv \frac{1}{\sigma}\bigg(\sigma(c_{\theta_Z}\times c_{\theta_f}> 0)
-\sigma(c_{\theta_Z} \times c_{\theta_f}< 0)\bigg).
\end{eqnarray}
Similarly $A_{xz}$ and $A_{yz}$ related to $T_{xz}$ and $T_{yz}$ are modified as
\begin{eqnarray}
\wtil{A}_{xz}\equiv \frac{1}{\sigma}\bigg(\sigma(c_{\theta_Z}\times c_{\theta_f}c_{\phi_f}> 0)
-\sigma(c_{\theta_Z} \times c_{\theta_f}c_{\phi_f}< 0)\bigg),\nonumber\\
\wtil{A}_{yz}\equiv \frac{1}{\sigma}\bigg(\sigma(c_{\theta_Z}\times c_{\theta_f}s_{\phi_f}> 0)
-\sigma(c_{\theta_Z} \times c_{\theta_f}s_{\phi_f}< 0)\bigg).
\end{eqnarray}
Redefining these asymmetries increases the total number of the non-vanishing 
observables to put simultaneous limit on the anomalous coupling and we expect 
limits  tighter than reported earlier in Ref.~\cite{Rahaman:2016pqj}. 

\begin{figure*}
\centering
\includegraphics[scale=0.6]{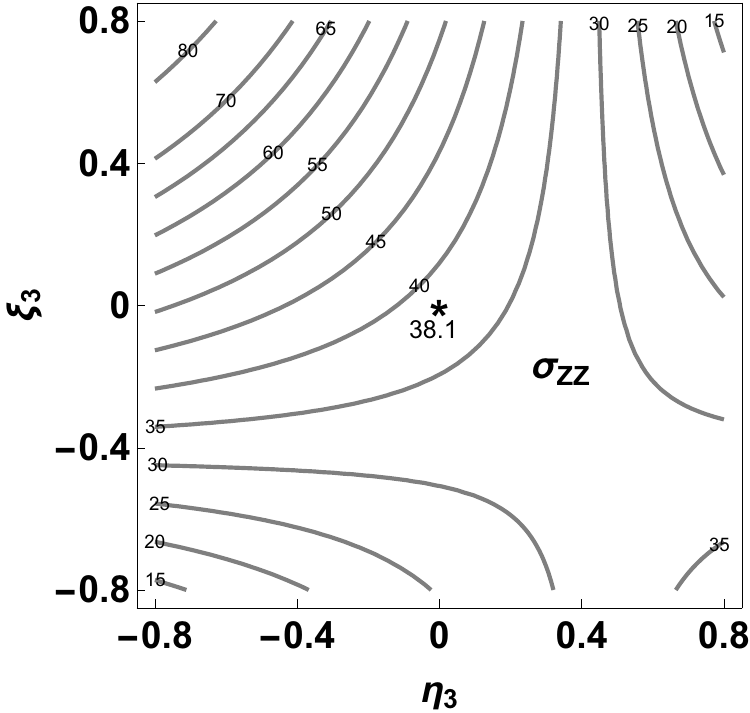}
\includegraphics[scale=0.6]{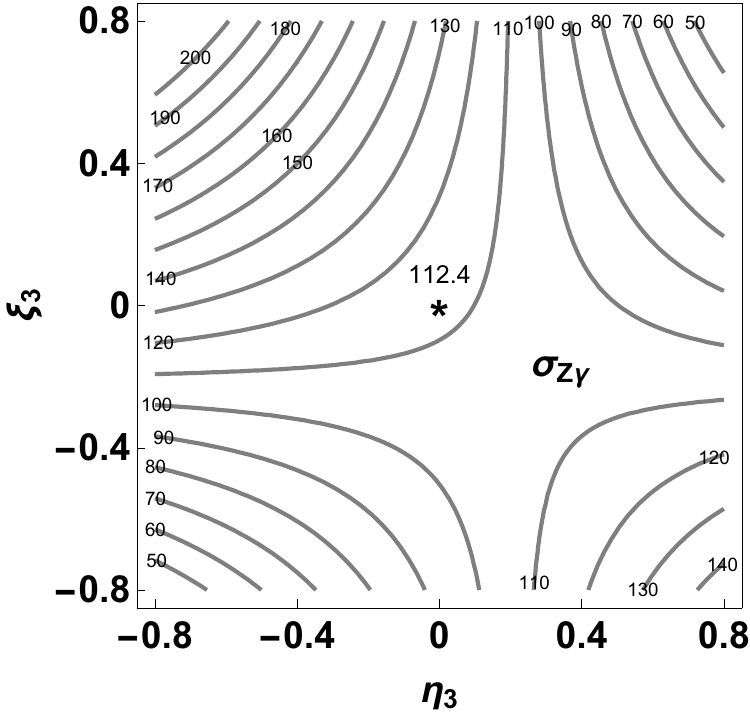}
\caption{\label{fig:beampol_cross-sections} 
The SM cross section (in fb) for the process $e^+e^-\to ZZ/Z\gamma$ as a function of
longitudinal beam polarizations $\eta_3$ (for $e^-$) and $\xi_3$ (for $e^+$)
at $\sqrt{s}=500$ GeV }
\end{figure*}

The total cross section (or total number of events) of a process plays an
important role determining the sensitivity and the limits on the anomalous 
couplings. A tighter limit on the anomalous couplings can be obtained if the 
cross section can be enhanced. Beam polarization can enhance the 
cross section and hence  it is important to see how it depends on beam 
polarization. Fig.~\ref{fig:beampol_cross-sections} shows the dependence of the 
cross sections $\sigma_{ZZ}$  and  $\sigma_{Z\gamma}$  on the longitudinal beam 
polarizations $\eta_3$ and $\xi_3$ at $\sqrt{s}=500$ GeV. The asterisk mark 
on the middle of the plots represents the unpolarized case. We notice that
the cross section in  the two processes are larger for negative value of $\eta_3$ 
and positive value of $\xi_3$. The sensitivity on the cross section is expected to be 
high in the left-top corner of the $\eta_3-\xi_3$ plane. This would convince 
us to set beam polarizations at the left-top corner for analysis. But the
cross section is not the only observable, the asymmetries have different 
behaviour on beam polarizations. For example, $A_x$ peaks at the right-bottom
corner, i.e. we have an opposite behaviour compared to cross section, while $A_z$ has
a similar dependence as cross section on the beam polarizations in both the 
processes. Processes involving $W^\pm$ are also expected to have higher 
cross section  at the left-top corner of $\eta_3-\xi_3$ plane as $W$ couple 
to the left chiral electron. Anomalous couplings are expected to change the 
dependence of all the observables, including the cross section, on the beam 
polarizations. To explore this we study the effect of beam polarizations on 
sensitivity of cross section and other observables to anomalous couplings 
in the next section.

%%%%%%%%%%%%%%%%%%%%%%%%%%%%%%%%%%%%%%%%%%%%%%%%%%%%%%%%%%%%%%%%%%%%%%%%%%%%%%
\section{Sensitivity, likelihood and the choice of beam polarizations}
\label{sec:sensitivity_likelihood}
%-----------------------------------------------------------------------------
The sensitivity of an observables ${\cal O}$ depending on anomalous couplings 
$\vec{f}$ with a given beam polarizations $\eta_3$ and $\xi_3$ is given by
\begin{equation}
{\cal S}({\cal O}(\vec{f},\eta_3,\xi_3))=\dfrac{|{\cal O}(\vec{f},\eta_3,\xi_3)
	-{\cal O}(\vec{0},\eta_3, \xi_3)|}{|\delta{\cal O}(\eta_3, \xi_3)|} \ \ ,
\end{equation}
where $\delta{\cal O}=\sqrt{(\delta{\cal O}_{stat.})^2+
(\delta{\cal O}_{sys.})^2}$ is the estimated error in ${\cal O}$. 
The estimated error to cross section would be
\begin{equation}
\delta\sigma(\eta_3, \xi_3)=\sqrt{\frac{\sigma(\eta_3, \xi_3)}{{\cal L}} +
	 \epsilon_\sigma^2 \sigma(\eta_3, \xi_3)^2 } \ \ ,
\end{equation}
whereas the estimated error to the asymmetries would be
\begin{equation}
\delta A(\eta_3, \xi_3)=\sqrt{\frac{1-A(\eta_3, \xi_3)^2}
	{{\cal L}\sigma(\eta_3, \xi_3)} + \epsilon_A^2 } \ \ .
\end{equation}
Here ${\cal L}$ is the integrated luminosity, $\epsilon_\sigma$ and 
$\epsilon_A$ are the systematic fractional error in cross section and 
asymmetries, respectively. In these analyses we take ${\cal L}=100$ fb$^{-1}$,
$\epsilon_\sigma=0.02$ and $\epsilon_A=0.01$ as a benchmark. 
%%%%%%%%%%%%%%%%%%%%%%%%%%%%%%%%%%%%%%%%%%%%%%%%%%%%%%%%%%%%%%%%%%%%
\begin{figure*}
\centering
\includegraphics[scale=0.48]{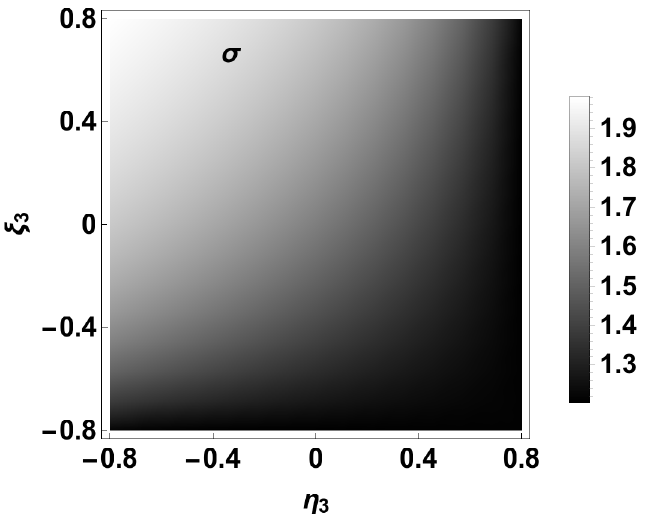}
\includegraphics[scale=0.48]{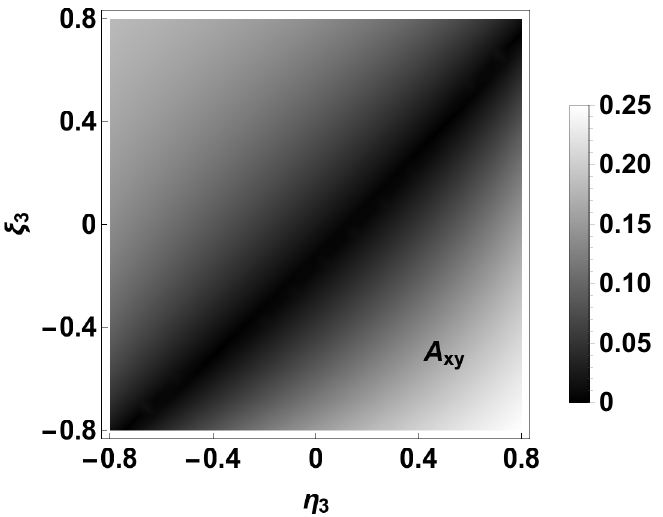}
\includegraphics[scale=0.48]{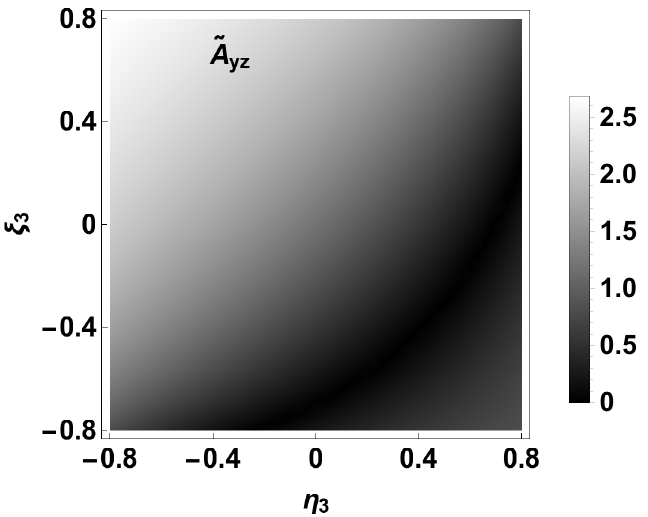}
\caption{\label{fig:sensitivity_zz}Effect of beam polarizations on sensitivity
of cross section $\sigma$, $A_{xy}$ and $\wtil{A}_{yz}$ in the process 
$e^+e^-\to ZZ$ for anomalous couplings $\vec{f}=\{+3,+3,+3,+3\}\times 10^{-3}$
at $\sqrt{s}=500$ GeV and ${\cal L}=100$ fb$^{-1}$}
\end{figure*}
%%%%%%%%%%%%%%%%%%%%%%%%%%%%%%%%%%%%%%%%%%%%%%%%%%%%%%%%%%%%%%%%%%%%
\begin{figure*}
\centering
\includegraphics[scale=0.47]{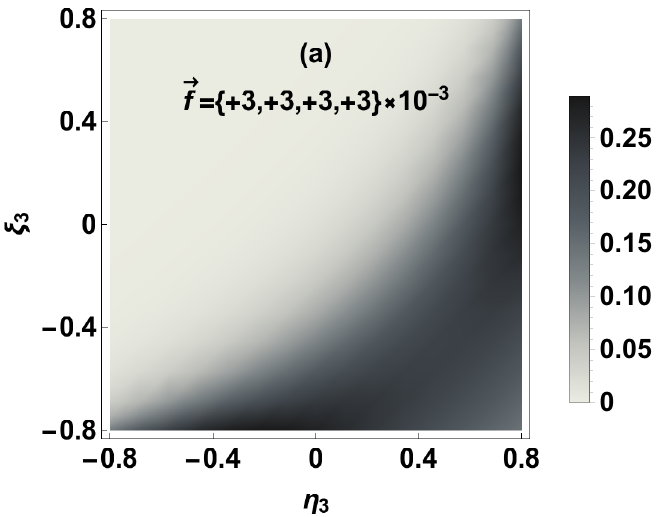}
\includegraphics[scale=0.47]{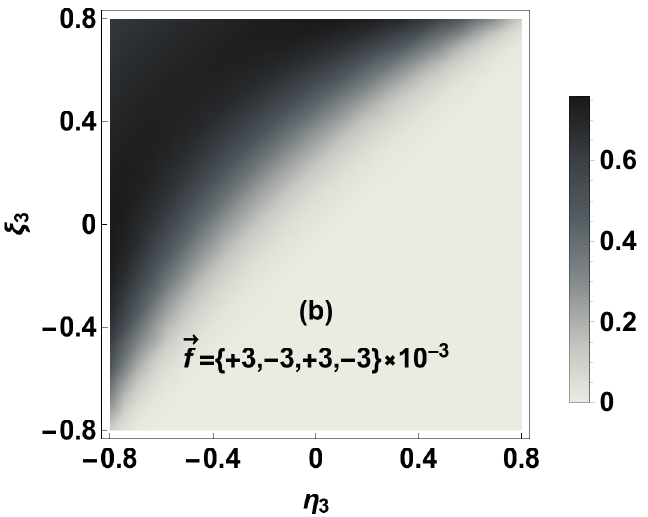}
\includegraphics[scale=0.47]{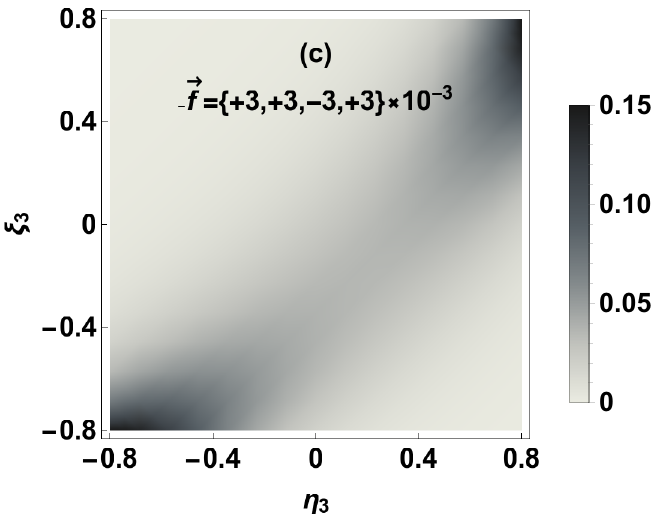}
\caption{\label{fig:x2_all_Obszz} Likelihood 
$\text{L}({\{\cal O}\},\vec{f};\eta_3,\xi_3)$ for three different benchmark 
anomalous couplings at $\sqrt{s}=500$ GeV and ${\cal L}=100$ fb$^{-1}$ in 
$ZZ$ process}
\end{figure*}
%%%%%%%%%%%%%%%%%%%%%%%%%%%%%%%%%%%%%%%%%%%%%%%%%%%%%%%%%%%%%%%%%%%%

We study the sensitivity of all the observables and their dependence on the beam
polarizations at $\sqrt{s}=500$ GeV. Choosing a benchmark value for the anomalous
couplings to be
$$\vec{f}=\{f_4^\gamma ,f_4^Z, f_5^\gamma, f_5^Z \} =\{+3,+3,+3,+3\}\times 
10^{-3} \ \ ,$$
we show the sensitivities for $\sigma$, $A_{xy}$ and $\wtil{A}_{yz}$ in 
Fig.~\ref{fig:sensitivity_zz} as a function of beam polarizations. The 
sensitivities for cross section and $\wtil{A}_{yz}$ peak at the left-top corner
of the plots. For $A_{xy}$ sensitivity peak at the right-bottom corner, it is 
not much smaller in the left-top corner either. The sensitivities of all other 
asymmetries (not shown here) except $\wtil{A}_{z}$ peaks at the left-top corner
although the exact dependence on the beam polarization may differ. Thus, the 
combined sensitivity of all the observables is high on the left-top corner
of the polarization plane making $(\eta_3,\xi_3)=(-0.8,+0.8)$ the best-choice
for the chosen benchmark coupling. This best-choice, however, strongly depends
upon the values of the anomalous couplings. 
We note that the best-choice of the beam polarization is mainly decided by the
behaviour of the cross section because most of the asymmetries also have 
similar dependences on the beam polarizations. This, however, does not mean that
the cross section provides a best sensitivity or the limits. For example, in
Fig.~\ref{fig:sensitivity_zz} we can see that $\wtil{A}_{yz}$ has a better
sensitivity than the cross section.
For the $Z\gamma$ process with  the benchmark point 
$$\vec{h}=\{h_1^\gamma,h_1^Z,h_3^\gamma,h_3^Z\}=\{+3,+3,+3,+3\}\times10^{-3}$$
one obtains similar conclusions: the sensitivities of all observables peak at left-top corner of $\eta_3-\xi_3$ plane (not shown) except for $\wtil{A}_{z}$.

For a complete analysis we need to use all the observables simultaneously.
To this end we define a likelihood function considering the set of all the 
observables depending on the anomalous coupling $\vec{f}$ as 
\begin{eqnarray}\label{eq:likelihood_zz}
\text{L}({\{\cal O}\},\vec{f};\eta_3,\xi_3)=
\exp{\bigg[-\frac{1}{2}\sum_{i} {\cal S}({\cal O}_i(\vec{f},\eta_3, 
\xi_3))^2 \bigg]} \ \ ,
\end{eqnarray}
$i$ runs over the set of observables in a process. Maximum sensitivity of 
observables requires the likelihood to be minimum. The likelihood defined here
is proportional to the $p$-value and hence the best-choice of beam polarizations 
comes from the {\em minimum} likelihood or maximum distinguishability.

The beam polarization dependence of the likelihood for the $ZZ$ process at the
above chosen anomalous couplings is given in Fig.~\ref{fig:x2_all_Obszz}(a).
The minimum of the likelihood falls in the left-top corner of the 
$\eta_3-\xi_3$ plane as expected as most of the observables has higher 
sensitivity at this corner. For different anomalous couplings the minimum 
likelihood changes its position in the $\eta_3-\xi_3$ plane.
We have checked the likelihood for $16$ different corners of 
$$\vec{f}_{\pm\pm\pm\pm}=\{\pm3,\pm3,\pm3,\pm3\}\times 10^{-3}$$ 
and they have different dependences on  $\eta_3 , \xi_3$. Here we present the
likelihood for three different choices of the anomalous couplings in 
Fig.~\ref{fig:x2_all_Obszz}. In Fig.~\ref{fig:x2_all_Obszz}(b), the minimum of
the likelihood falls in the right-bottom corner where most of the observables have
higher sensitivity. In Fig.~\ref{fig:x2_all_Obszz}(c) low likelihood falls in 
both  diagonal corners in the $\eta_3-\xi_3$ plane. This is because some of the 
observables prefer the left-top corner, while others prefer the right-bottom corner 
of the polarization plane for higher sensitivity. We have a similar behaviour
for the likelihood in the $Z\gamma$ process.

As the anomalous couplings change, the minimum likelihood region changes 
accordingly and hence the best-choice of beam polarizations. So the best-choice
for the beam polarizations depends on the new physics in the process. If one
knows the new physics one could tune the beam polarizations to have the best 
sensitivity for the analysis. But in order to have a suitable choice of beam 
polarizations irrespective of the possible new physics one needs to minimize 
the likelihood averaged over all the anomalous couplings. 
The likelihood 
function averaged  over a volume in parameter space $V_{\vec{f}}$ would be 
defined as
\begin{equation}\label{eq:average_likelihood_zz}
L(V_{\vec{f}},{\{\cal O}\};\eta_3,\xi_3)=\int_{V_{\vec{f}}}
\text{L}({\{\cal O}\},\vec{f};\eta_3,\xi_3) d\vec{f}.
\end{equation}
This quantity is nothing but the {\em weighted volume} of the parameter space that is
statistically consistent with the SM. The size of this weighted volume determines
the limits on the parameters. The beam polarizations with the minimum averaged 
likelihood (or minimum weighted volume) is expected to be the average best 
choice for any new physics in the process. For numerical analysis we
choose the volume to be a hypercube in the $4$ dimensional parameter space 
with sides equal to $2\times 0.05$ (much larger than the 
available limits on them) in both the processes. The contribution to the average
likelihood from the region outside this volume is negligible. 
%%%%%%%%%%%%%%%%%%%%%%%%%%%%%%%%%%%%%%%%%%%%%%%%%%
\begin{table*}
\centering
\caption{\label{tab:coupling_limit_mcmc_ZZ} List of simultaneous limits on 
anomalous couplings obtained for $\sqrt{s}=500$ GeV and $L=100$ fb$^{-1}$
for different $\eta_3$ and $\xi_3$ from MCMC in $ZZ$ process}
\renewcommand{\arraystretch}{1.5}
\begin{tabular*}{\textwidth}{@{\extracolsep{\fill}}lllllllllllllll@{}}\hline
\multicolumn{1}{c}{\begin{tabular}{l} Beam\\polarizations \end{tabular}} &
\multicolumn{8}{c}{   Limits on couplings ($ 10^{-3}$)}\\ \hline
\multicolumn{1}{c}{}  & 	\multicolumn{2}{c}{$f_4^\gamma $} & 	\multicolumn{2}{c}{$f_4^Z$} & 	\multicolumn{2}{c}{$f_5^\gamma$} & 	\multicolumn{2}{c}{$f_5^Z$}\\\hline
~~~~~$(\eta_3,\xi_3)$ & $68~\%$ & $95~\%$ & $68~\%$ & $95~\%$ & $68~\%$ & $95~\%$ & $68~\%$ & $95~\%$ & Comments \\ \hline
$-0.80,+0.80$ & $_{-9.3}^{+7.3} $& $_{-12.0}^{+13.0} $& $_{-14.0}^{+15.0} $& $_{-19.0}^{+18.0}  $& $\pm 7.3 $& $\pm 13.0  $ & $\pm 11.0 $& $_{-18.0}^{+19.0}  $ &\\
$-0.40,+0.40$ & $\pm3.1 $& $_{-5.7}^{+5.8} $& $\pm 4.4 $& $_{-8.4}^{+8.2}  $& $\pm3.3 $& $_{-6.2}^{+6.3}  $& $_{-5.2}^{+4.5}  $& $_{-8.5}^{+9.3}  $&\\ \hline
~~~$0.00,~~~0.00$ & $\pm1.7 $& $\pm3.3 $& $\pm2.5 $& $\pm4.8 $ & $\pm1.9 $& $_{-3.6}^{+3.7}$ & $_{-2.7}^{+2.3}$& $_{-4.6}^{+5.1} $& Unpolarized point\\ 
$+0.09,-0.10$ & $\pm 1.7$& $\pm 3.2 $& $\pm2.4 $& $_{-4.6}^{+4.7}  $& $\pm1.8 $&
$_{-3.4}^{+3.5} $& $_{-2.6}^{+2.2} $& $_{-4.5}^{+4.9} $ & $P_{Z\gamma}$, 
best-choice for $Z\gamma$ \\
$+0.12,-0.12$ & $\pm 1.6$& $\pm 3.1$& $\pm2.4 $& $\pm4.7 $& $\pm1.8 $&
$_{-3.4}^{+3.5} $& $_{-2.6}^{+2.2} $& $_{-4.5}^{+5.0} $ & $P_{best}$, combined best-choice \\ 
$+0.16,-0.16$ & $\pm 1.6$& $\pm3.1 $& $\pm2.4 $& $\pm4.7 $& $\pm1.8 $&
$_{-3.4}^{+3.5} $& $_{-2.7}^{+2.3} $& $_{-4.5}^{+5.1} $ & $P_{ZZ}$, best-choice for $ZZ$\\ \hline
$+0.40,-0.40$ & $\pm1.9 $& $\pm3.7$& $\pm3.2 $& $_{-6.2}^{+6.1} $& $\pm2.1 $& $_{-4.1}^{+4.0} $& $_{-3.7}^{+3.1} $& $_{-6.0}^{+6.7} $&\\
$+0.80,-0.80$ & $_{-6.2}^{+5.3} $& $_{-9.3}^{+9.8} $& $_{-12.0}^{+9.7} $& $_{-17.0}^{+18.0} $& $\pm5.4 $& $_{-9.9}^{+9.5} $& $\pm9.9 $& $_{-18.0}^{+17.0} $&\\\hline
\end{tabular*}
\end{table*}
%%%%%%%%%%%%%%%%%%%%%%%%%%%%%%%%%%%%%%%%%%%%%%%%%%
\begin{figure}
\centering
\includegraphics[scale=0.7]{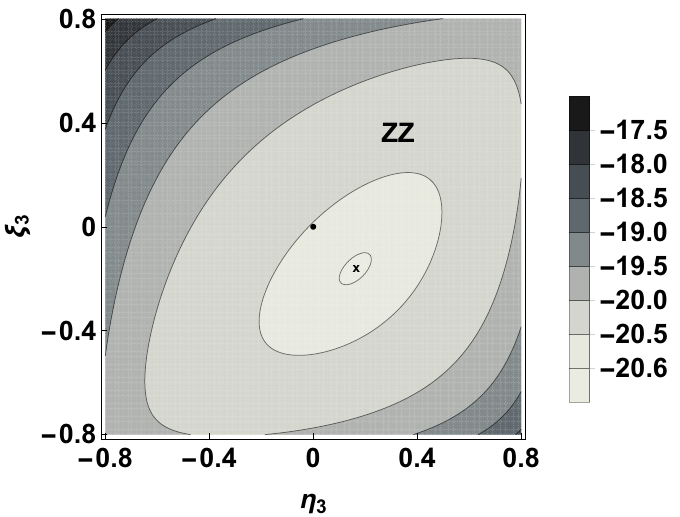}
\caption{\label{fig:x2_all_ObsZZ_Int}The log of average likelihood, 
$\log[ L(V_{\vec{f}},{\{\cal O}\};\eta_3,\xi_3)]$ as a function of beam
polarization is shown for the $ZZ$ process at $\sqrt{s}=500$ GeV and
${\cal L}=100$ fb$^{-1}$. The dot at the centre is the ($0,0$) point, while 
cross mark at $P_{ZZ}=(+0.16,-0.16)$ is the minimum likelihood point and 
hence the best-choice of beam polarizations for $ZZ$ process}
\end{figure}
%%%%%%%%%%%%%%%%%%%%%%%%%%%%%%%%%%%%%%%%%%%%%%%%%%

The average likelihood $L(V_{\vec{f}},{\{\cal O}\};\eta_3,\xi_3)$  in the $ZZ$
process as a function of  beam polarization is shown in 
Fig.~\ref{fig:x2_all_ObsZZ_Int} on $\log$-scale. The dot on the middle of the 
plot represents the unpolarized case and the cross mark at 
$P_{ZZ}=( +0.16,-0.16)$ represents the minimum averaged likelihood point i.e.,
the best-choice of beam polarizations. 
%This choice is not much far away from 
%the unpolarized case.
The unpolarized point, the best point and the points within 
two central contour 
in Fig.~\ref{fig:x2_all_ObsZZ_Int} have the same order of average likelihood 
and expected to give similar limits on anomalous couplings. The polarization
point from darker contours corresponds to larger values of average likelihood
and expected to give relatively looser limits on anomalous couplings.
To explore this we estimate simultaneous limits using 
Markov-Chain-Monte-Carlo (MCMC) method at $P_{ZZ}$, unpolarized beam and few 
other benchmark choice of beam polarizations. The limits thus obtained on the 
anomalous couplings for the  $ZZ$ process are listed in 
Table~\ref{tab:coupling_limit_mcmc_ZZ}. We note that the limits for the 
best-choice of polarizations ($P_{ZZ}$) are best but comparable to other 
nearby benchmark beam polarization including the unpolarized beams.
This is due to the fact that the average likelihood is comparable for these 
cases. Further, the limits for $(+0.4,-0.4)$ and $(-0.4,+0.4)$
are increasingly bad as these points correspond to the third and fourth contours,
i.e., we have increasingly larger average likelihood. The point $(-0.8,+0.8)$ has the
largest average likelihood and the corresponding limits are the worst in 
Table~\ref{tab:coupling_limit_mcmc_ZZ}. 
We also note that the limits for the unpolarized case in
Table~\ref{tab:coupling_limit_mcmc_ZZ} are better than the ones reported in
Ref.~\cite{Rahaman:2016pqj}, when adjusted for the systematic errors. This
improvement  here is due to the inclusion of three new non-vanishing asymmetries
$\wtil{A}_z$, $\wtil{A}_{xz}$ and $\wtil{A}_{yz}$. Of these, $\wtil{A}_{xz}$ has 
linear dependence on $f_5^{\gamma,Z}$ with larger sensitivity to $f_5^Z$ leading
to about $30~\%$ improvement in the limit. Similarly, the $CP$-odd asymmetry
$\wtil{A}_{yz}$ has a linear dependence on $f_4^{\gamma,Z}$ with larger
sensitivity to $f_4^Z$ and this again leads to about  $30~\%$ improvement in the
corresponding limit. The asymmetry $\wtil{A}_z$ has a quadratic dependence on all
four parameters and has too poor sensitivity for all of them to be useful.

\begin{figure}
\centering
\includegraphics[scale=0.7]{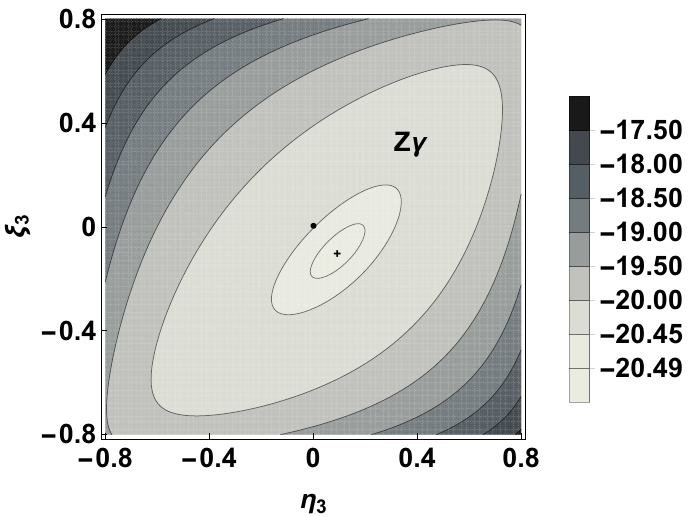}
\caption{\label{fig:x2_all_ObsZA_Int} 
Same as Fig.~\ref{fig:x2_all_ObsZZ_Int} but for the $Z\gamma$ process.
The plus mark at $P_{Z\gamma}=(+0.09,-0.10)$ is the lowest likelihood point 
and hence the best-choice of beam polarizations for $Z\gamma$ process}
\end{figure}

We do a similar analysis for the $Z\gamma$ process. The average likelihood  
$L(V_{\vec{h}},{\{\cal O}\};\eta_3,\xi_3)$ is shown in 
Fig.~\ref{fig:x2_all_ObsZA_Int} on log-scale.
Here also the dot on the middle of the plot is for unpolarized case while the
plus mark at $P_{Z\gamma}=( +0.09,-0.10)$ is for the minimum averaged likelihood 
and hence the best-choice of beam polarizations. The corresponding simultaneous 
limits on the anomalous couplings $h_i$ are presented in 
Table~\ref{tab:coupling_limit_mcmc_ZA}.Again we notice that the limits obtained
for the best-choice of the beam polarizations $P_{Z\gamma}$ are tighter than any 
other point on the polarization plane,  yet comparable to the nearby polarization
points within the two central contours in Fig.~\ref{fig:x2_all_ObsZA_Int}, including the unpolarized point. This again is due to the comparable
values of the averaged likelihood of the two
central contours containing $P_{Z\gamma}$ and the unpolarized point. The limits at the points $(+0.4,-0.4)$ 
and $(-0.4,+0.4)$ are  worse as they fall in the fourth and fifth contour containing much 
larger likelihood values. Like the $ZZ$ case the point $(-0.8,+0.8)$ has the
largest average likelihood and the corresponding limits are the worst. 
The simultaneous limits for the unpolarized case (also the $P_{Z\gamma}$) turns out
to be much better than the ones reported in Ref.~\cite{Rahaman:2016pqj} for 
$h_{1,3}^\gamma$ due to the inclusions of new asymmetries in the present
analysis. The $CP$-odd asymmetry $\wtil{A}_{yz}$ has linear dependence on
$h_1^{\gamma,Z}$ with a large sensitivity towards $h_1^\gamma$ leading to an
improvement in the corresponding limit by a factor of two compare to earlier 
report when adjusted for systematic errors. The limit on $h_3^\gamma$ improves 
by a factor of $3$ owing to the asymmetry $\wtil{A}_{xz}$. The limits on 
$h_{1,3}^Z$ remain comparable.

\begin{table*}
\centering
\caption{\label{tab:coupling_limit_mcmc_ZA} List of simultaneous limits on 
anomalous couplings obtained for $\sqrt{s}=500$ GeV and $L=100$ fb$^{-1}$
for different $\eta_3$ and $\xi_3$ from MCMC in $Z\gamma$ process}
\renewcommand{\arraystretch}{1.5}
\begin{tabular*}{\textwidth}{@{\extracolsep{\fill}}lllllllllllllll@{}}\hline
\multicolumn{1}{c}{\begin{tabular}{l} Beam\\polarizations \end{tabular}} &
\multicolumn{8}{c}{   Limits on couplings ($ 10^{-3}$)}\\ \hline
\multicolumn{1}{c}{}  & 	\multicolumn{2}{c}{$h_1^\gamma $} & 	\multicolumn{2}{c}{$h_1^Z$} & 	\multicolumn{2}{c}{$h_3^\gamma$} & 	\multicolumn{2}{c}{$h_3^Z$}\\\hline
~~~~~$(\eta_3,\xi_3)$ & $68~\%$ & $95~\%$ & $68~\%$ & $95~\%$ & $68~\%$ & $95~\%$ & $68~\%$ & $95~\%$ & Comments \\ \hline
$-0.80,+0.80$ & $_{-9.3}^{+7.7} $& $\pm 13.0$& $\pm 11.0$& $_{-19.0}^{+18.0}  $& $\pm 7.5 $& $\pm 13.0  $ & $\pm 11.0 $& $\pm 19.0$ &\\
$-0.40,+0.40$ & $\pm3.9 $& $_{-7.5}^{+7.4} $& $\pm 6.5 $& $\pm 12.0$& $_{-3.7}^{+4.4}$& $_{-8.0}^{+7.1}  $& $\pm 6.6$& $_{-12.0}^{+13.0}  $ &\\ \hline
~~~$0.00,~~~0.00$ & $\pm1.6 $& $\pm3.1 $ & $\pm3.7 $& $_{-7.0}^{+7.1}$& $_{-1.4}^{+1.6} $& $_{-3.0}^{+2.8} $ & $\pm 3.6$& $\pm 7.1 $ & Unpolarized point\\ 
$+0.09,-0.10$ & $\pm 1.5$& $\pm2.9 $& $\pm3.6 $& $\pm7.0 $& $_{-1.3}^{+1.4}$&
$_{-2.8}^{+2.6} $& $\pm 3.6$& $_{-7.1}^{+7.0}$& $P_{Z\gamma}$, best-choice for $Z\gamma$\\
$+0.12,-0.12$ & $\pm 1.5$& $\pm2.9 $& $\pm3.7 $& $\pm7.1 $& $\pm 1.4$&
$_{-2.8}^{+2.6} $& $\pm 3.6$& $\pm 7.1$ & $P_{best}$, combined best-choice\\
$+0.16,-0.16$ & $\pm 1.5$& $\pm3.0 $& $\pm3.7 $& $_{-7.3}^{+7.2} $& $_{-1.3}^{+1.5}
$& $_{-2.8}^{+2.6}$ & $\pm 3.7$& $_{-7.3}^{+7.1}$ & $P_{ZZ}$, best-choice for $ZZ$\\\hline
$+0.40,-0.40$ & $\pm2.4 $& $\pm4.6$& $\pm5.2 $& $\pm 10.0$& $_{-2.2}^{+2.5} $& $_{-4.7}^{+4.3} $& $\pm 5.2 $& $\pm 10.0 $&\\
$+0.80,-0.80$ & $\pm 5.8$& $_{-9.9}^{+10.0} $& $_{-13.0}^{+11.0} $& $_{-18.0}^{+19.0} $& $_{-7.2}^{+5.8} $& $_{-9.7}^{+10.0} $& $_{-15.0}^{+13.0}$& $_{-18.0}^{+19.0} $ &\\\hline
\end{tabular*}
\end{table*}

\begin{figure}
\centering
\includegraphics[scale=0.7]{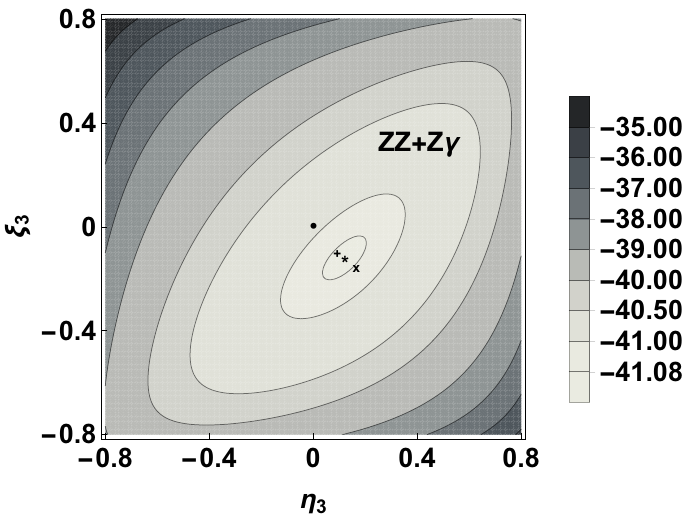}
\caption{\label{fig:x2_all_ObsZZZA_Int}The log of average likelihood, 
$\log[L(V_{\{\vec{f},\vec{h}\}},{\{\cal O}\};\eta_3,\xi_3)]$, is shown 
considering both the processes $ZZ$ and $Z\gamma$ at $\sqrt{s}=500$ GeV, 
${\cal L}=100$ fb$^{-1}$. The asterisk mark at $P_{best}=(+0.12,-0.12)$ is the
combined best-choice for beam polarizations while the other points are for 
$ZZ$ ({\em cross mark}) and $Z\gamma$ ({\em plus mark}) }
\end{figure}

The combined analysis of the processes $ZZ$ and $Z\gamma$ is expected to change
the best-choice of beam polarizations and limits accordingly. For the average 
likelihood for these two processes the volume, in which one should  average, 
will change to $V_{\vec{f}/\vec{h}} \to V_{\vec{F}}$, where 
$\vec{F}=\{\vec{f},\vec{h}\}$ and observables from both  processes should be 
added to the likelihood defined in Eq.~\ref{eq:likelihood_zz}. The combined averaged
likelihood showing dependence on the beam polarizations for the two processes 
considered here is shown in Fig.~\ref{fig:x2_all_ObsZZZA_Int}. The dot on the 
middle of the plot is for the unpolarized case and asterisk mark at 
$P_{best}=(+0.12,-0.12)$ is the combined best-choice of beam polarizations. 
Other points are due to $P_{ZZ}$ and $P_{Z\gamma}$. The combined best-choice 
point sits in between $P_{ZZ}$ and $P_{Z\gamma}$. The limits, presented in 
Table~\ref{tab:coupling_limit_mcmc_ZZ} and~\ref{tab:coupling_limit_mcmc_ZA}, 
at the combined best-choice of the beam polarizations are slightly weaker than 
the limit at the best-choice points but comparable in both  processes as 
expected. Thus the combined best-choice can be a good benchmark beam 
polarizations for the process $ZZ$ and $Z\gamma$ to study at ILC.

The best-choice of beam polarizations, obtained here, depends 
on the size of the estimated error of the observables and hence on the 
systematics $\epsilon_\sigma$ and $\epsilon_A$. Numerical analysis shows that 
the best-choice points, for both  processes separately and combined, move away 
from the unpolarized point along the cross diagonal axis towards the right-bottom corner on the $\eta_3 - \xi_3$ plane when $\epsilon_\sigma$ or
$\epsilon_A$ or both are increased. For example, if we double $\epsilon_\sigma$
and $\epsilon_A$ both, i.e. we take $\epsilon_\sigma=0.04$ and $\epsilon_A=0.02$, 
the best-choice points $P_{ZZ}$, $P_{Z\gamma}$ and $P_{best}$ become 
($+0.20,-0.20$),  ($+0.13,-0.12) $ and ($+0.17,-0.16 $), respectively.
On the other hand the best-choice points move towards the unpolarized point as
the systematics are decreased. For example, when the systematics are reduced by
$1/2$, i.e. for $\epsilon_\sigma=0.01$ and $\epsilon_A=0.005$, the best-choice 
points for $ZZ$, $Z\gamma$ and for combined process  move to ($+0.15,-0.15 $),
($+0.08,-0.08 $) and ($+0.11,-0.11 $), respectively. However, the best-choice 
points do not move further closer to the unpolarized point when the size of 
systematics becomes smaller than the statistical one.

Similar analysis as presented in Fig.~\ref{fig:x2_all_ObsZZZA_Int}
can be done by combining many processes, as one should 
do, to choose a suitable beam polarizations at ILC. For many processes
(e.g., we can include $WW$ production with anomalous couplings among charged 
gauge boson) with different couplings, the volume in which one should do the 
average will change to $V_{\vec{f}/\vec{h}}= V_{\vec{F}}$, where $\vec{F}$ would be the
set of all couplings for all the processes considered. The set of observables
$\{\mathcal{O}\}$ would include all the relevant observables from all the
processes combined in the expression for the likelihood.
%%%%%%%%%%%%%%%%%%%%%%%%%%%%%%%%%%%%%%%%%%%%%%%%%%%%%%%%%%%%%%%%%%%%%%%%%%%%%%%%%%%%%%%
\section{Conclusion}\label{sec:conclusion}
To summarize, we aim to find the best-choice of beam polarization for an
$e^+e^-$ collider to probe the anomalous couplings in the neutral gauge-boson
sector in the $ZZ$ and $Z\gamma$ processes. 
We study the effects of beam polarization on polarization asymmetries and
corresponding sensitivities towards anomalous couplings. Using the {\em minimum
averaged} likelihood, we find the best-choice of the beam polarization for the
two processes and also the combined best-choice. Here the list of observables 
includes the cross section along with eight polarization asymmetries for 
the $Z$ boson. Simultaneous limits on
anomalous couplings were obtained using the MCMC method for a set of benchmark 
beam polarization including the best-choices and they are listed in 
Tables~\ref{tab:coupling_limit_mcmc_ZZ} and \ref{tab:coupling_limit_mcmc_ZA}.
The limits obtained for the unpolarized case are better than the ones
reported in Ref.~\cite{Rahaman:2016pqj}. This is because the present analysis
includes three new observables $\wtil{A}_z$, $\wtil{A}_{xz}$ and
$\wtil{A}_{yz}$. These new asymmetries yield better limits on $f^Z_{4,5}$
and $h^\gamma_{1,3}$, while we have comparable (yet better) limits on $f^\gamma_{4,5}$
and $h^Z_{1,3}$. Comparing the limits for various benchmark beam polarizations
from Tables~\ref{tab:coupling_limit_mcmc_ZZ} and
\ref{tab:coupling_limit_mcmc_ZA}, we find that all three {\em best} beam 
polarization choices yield comparable limits and they are comparable to the 
unpolarized case as well. Thus, as far as anomalous couplings in the neutral
gauge-boson sector are concerned, unpolarized beams perform as good as the 
{\em best} choices. This conclusion, however, can change if one includes more
or a different set of observables in the analysis or add more processes to the
analysis. For example, processes involving $W^\pm$ or a Higgs boson may have a
different preference for the beam polarization. 

Considering the physics impact and the cost of beam polarizations at
ILC one may chose the unpolarized beams for the first run, at least for the two
processes studied here. But as we infer, a detailed global analysis is required
involving other processes as well to conclude this. We further note that the
case of transverse beam polarization is not addressed here and conclusions 
may differ in that case.

%%%%%%%%%%%%%%%%%%%%%%%%%%%%%%%%%%%%%%%%
\noindent \textbf{Acknowledgements:} R.~R. thanks Department of Science 
and Technology, Government of India for support through DST-INSPIRE Fellowship 
for doctoral program, INSPIRE CODE IF140075, 2014.  

%%%%%%%%%%%%%%%%%%%%%%%%%%%%%%%%%%%%%%%%%%%%%%%%%%%%%%%%%%%%%%%%%%%%%%%%%
\fontsize{9}{10}\selectfont
\bibliography{References/zpol-refs,References/ATGC_VecPol,References/Bibliography_nag,References/Bibliography0}
\bibliographystyle{utphys}
%%%%%%%%%%%%%%%%%%%%%%%%%%%%%%%%%%%%%%%%%%%%%%%%%%%%%%%%%%%%%%%%%%%%%%%%%%%
\end{document}